\documentclass{raa}
\usepackage{graphicx,times}
\usepackage{natbib}
\usepackage{amssymb,amsmath}
\bibpunct{(}{)}{;}{a}{}{,}

\usepackage{dcolumn}
\usepackage{hyperref}
\hypersetup{pdftitle = The title of my PDF, pdfauthor = My name, pdfsubject= The subject, pdfkeywords = keyword1 keyword2 keyword3} 
\hypersetup{colorlinks = true, linkcolor = green, anchorcolor = red, citecolor = blue, filecolor = red, pagecolor = red, urlcolor = red}

\newcommand{\hMsun}{{\ifmmode{h^{-1}{\rm {M_{\odot}}}}\else{$h^{-1}{\rm{M_{\odot}}}$}\fi}}

\begin{document}

   \title{HIKER: a halo-finding method based on kernel-shift algorithm}

   \volnopage{Vol.0 (20xx) No.0, 000--000}
   \setcounter{page}{1}

   \author{Shuangpeng Sun
      \inst{1,2}
   \and Shihong Liao
      \inst{1}
   \and Qi Guo
      \inst{1,2}
   \and  Qiao Wang
      \inst{1,2}
   \and Liang Gao
      \inst{1,2,3}
   }

   \institute{Key Laboratory of Computational Astrophysics, National Astronomical Observatories, Chinese Academy of Sciences, Beijing 100101, China; {\it sunshp@nao.cas.cn}\\
        \and
             University of Chinese Academy of Sciences, Beijing 100049, China\\
        \and
             Institute for Computational Cosmology, Department of Physics, Durham University, Science Laboratories, South Road, Durham DH1 3LE, England\\
\vs \no
   {\small Received  20xx month day; accepted  20xx  month day}}

\abstract{ 
We introduce a new halo/subhalo finder, HIKER (a Halo fInder based on KERnel-shift algorithm), which takes advantage of a machine learning method -- the mean-shift algorithm combined with the Plummer kernel function, to effectively locate density peaks corresponding to halos/subhalos in density field. Based on these density peaks, dark matter halos are identified as spherical overdensity structures, and subhalos are bound substructures with boundaries at their tidal radius. By testing HIKER code with mock halos, we show that HIKER performs excellently in recovering input halo properties. Especially, HIKER has higher accuracy in locating halo/subhalo centres than most halo finders. With cosmological simulations, we further show that HIKER reproduces the abundance of dark matter halos and subhalos quite accurately, and the HIKER halo/subhalo mass functions and $V_\mathrm{max}$ functions are in good agreement with two widely used halo finders, SUBFIND and AHF.
}

   \authorrunning{S. Sun et al. }
   \titlerunning{HIKER halo finder}

   \maketitle

\section{Introduction}\label{sec:intro} 
Cosmological N-body simulations are one of the most crucial methods to study structure formation and evolution of the universe \citep[see e.g.][for reviews]{Frenk2012,Kuhlen2012}. To compare simulations with observations so that cosmological models can be constrained, a key step is to identify gravitationally bound structures (e.g. dark matter halos and subhalos) from simulation data. Especially, modern cosmological simulations with large simulated volume and high numerical resolution resolve plenty of structures and substructures spanning a wide range of masses/sizes. How to identify dark matter halos/subhalos efficiently and robustly from these large simulations is of great importance. In the past decades, many codes/methods have been developed to identify halos/subhalos from simulation snapshots (i.e. halo finders); see \citet{Knebe2011} and \citet{Onions2012} for reviews.

Usually, halo finders can be classified into two categories, spherical overdensity \citep[SO,][]{Press1974,Lacey1994} and friends-of-friends \citep[FOF,][]{Davis1985} method. SO halo finders first locate peaks in the density field, then a sphere centring each peak is grown out until the mean matter density within this sphere reaches a given threshold. Usually this threshold is expressed as an overdensity parameter, $\Delta$, measured with respect to the mean matter density or the cosmic critical density. The final sphere defines the boundary of a halo. Examples of SO halo finders include BDM \citep[Bound Density Maxima,][]{Klypin1997} and AHF \citep[Amiga Halo Finder,][]{Knollmann2009}. FOF halo finders group simulation particles whose pairwise distances are smaller than a given threshold, which is usually expressed as the linking length parameter, $b$, measured with respect to the mean interparticle distance \citep{More2011}. The centre of an FOF halo is usually defined as the position of the most bound particle. Examples of FOF halo finders include SUBFIND \citep{Springel2001} and Rockstar \citep{Behroozi2013}. 

Recently, machine learning algorithms have been widely adopted in the astronomical community, and they are becoming more and more useful as the astronomical data size grows more rapidly. For example, \citet{Hui2018} use support vector machine (SVM) to classify galaxies to study the relation with large-scale structures; \citet{Aragon-Calvo2019} uses deep convolutional neural network (CNN) to perform large-scale structures classifications; \citet{He2019} use deep neural networks to predict the formation of non-linear large-scale structures. We refer the reader to \cite{Baron2019} for a recent review of various applications of machine learning methods in astronomy. Unsupervised machine learning algorithms are widely used in clustering analysis, outlier detection, dimensionality reduction, etc. Especially, the concept of clustering analysis for data points, which tries to group objects so that the objects in the same group (or cluster) are more similar to each other than objects in other groups (or clusters), is fairly similar to halo finding in cosmological simulations. This motivates us to develop a new halo finder code based on clustering analysis algorithms.

The mean-shift algorithm \citep{Fukunaga1975,YizongCheng1995, Comaniciu2002} is a non-parametric cluster finding procedure, and it is popular in computer vision and image processing applications. It starts from an initial position, and then iteratively shifts to the local maximum (i.e. mode) of a density field with a shift vector computed from the average property of data points in the neighborhood. The mean-shift algorithm does not assume the shape of the distribution nor the number of clusters, and it is quite efficient to locate the local maxima of a density distribution. These advantages make it an attractive candidate method to identify halos from cosmological simulations. Especially, as we show in this paper, the mean-shift algorithm equipping with a physically motivated kernel function, the Plummer kernel, has another advantage in accurately identifying substructures from simulations. To stress the feature of our algorithm (i.e. the mean-shift algorithm with the Plummer kernel function), we use the name of ``kernel-shift" in this paper.

In this work, we introduce a new halo finder, HIKER (a Halo-fInding method based on KERnel-shift algorithm) to simultaneously identify dark matter halos and subhalos in cosmological N-body simulations. The organization of our paper is as follows. Section \ref{sec:algorithm} introduces the kernel-shift algorithm and other details of our halo finding procedures. We then apply HIKER to mock halo data as well as cosmological N-body simulations, and present our test results on field halos and subhalos in Section \ref{sec:results}. Section \ref{sec:conclusions} presents our conclusions and discussions.

\section{HIKER algorithm}\label{sec:algorithm} 
The general procedures of the HIKER algorithm can be described  as follows: (i) Identifying density peaks with the kernel-shift algorithm. (ii) For each density peak, a sphere is grown around its centre until the enclosed mean density is 200 times the critical density. Halos and subhalos are determined according to their geometrical relations. (iii) For each subhalo, we determine its tidal radius and apply for an unbinding procedure to exclude unbound particles.

\subsection{Mean-shift algorithm}\label{subsec:mean_shift}
The mean-shift algorithm is one of the clustering algorithms in the unsupervised machine learning category, which is very efficient and robust to locate density peaks of a density field. The algorithm was first presented in \citet{Fukunaga1975} and later generalized by introducing kernel functions in \citet{YizongCheng1995}.

To locate the local density maxima of a set of particles, the mean-shift algorithm starts from an initial guessing position, moves iteratively according to mean-shift vectors computed from the neighboring particles, until the convergence is reached. The key concept of the algorithm is the mean-shift vector. In the original mean-shift algorithm proposed by \citet{Fukunaga1975}, for a position $\bm{x}$, the mean-shift vector is computed as
\begin{equation}
 \bm{M}_b(\bm{x})=\frac{1}{N}\sum_{i=1}^{N}\left( \bm{x}_i-\bm{x} \right),
\end{equation}
where $\bm{x}_i$ is the coordinate of the $i$-th particle whose distance to position $\bm{x}$ is less than the bandwidth radius, $b$, and $N$ is the total number of such particles. Note that the bandwidth radius here is the single input parameter in the mean-shift algorithm. Physically, if a given set of particles have equal masses, the mean-shift vector is defined as the difference between position $\bm{x}$ and the centre-of-mass position of the particles within the bandwidth radius.

\citet{YizongCheng1995} later generalized the mean-shift vector by introducing a Kernel function, $K(y; b)$,
\begin{equation}\label{eq:mean_shift_vector}
    \bm{M}_b^K(\bm{x})=\frac{\sum_{i=1}^{N}K(\left| \bm{x}_i-\bm{x} \right|; b)\left( \bm{x}_i-\bm{x} \right)}{\sum_{i=1}^{N}K(\left| \bm{x}_i-\bm{x} \right|; b)},
\end{equation}
where $\left| \bm{x}_i-\bm{x} \right|$ is the distance from $\bm{x}$ to $\bm{x}_i$, and $N$ here denotes the total number of particles which have their $\left| \bm{x}_i-\bm{x} \right|$ smaller than an input radius $r_K$. Under such generalizations, the original mean-shift vector, $\bm{M}_b(\bm{x})$, can be regarded as $\bm{M}_b^K(\bm{x})$ computed with a flat kernel, i.e. $K(y; b)= 1$ if $y \leq b$, and $0$ otherwise. For each position, it can be moved to a new position with higher density according to the mean-shift vector, i.e. $\bm{x}_\mathrm{new} = \bm{x} + \bm{M}_b^K(\bm{x})$. Iteratively repeating the procedure leads to convergence once it reaches a local density maximum. For the detailed mathematical proof of the convergence for the mean-shift algorithm, see \citet{YizongCheng1995}.

In HIKER, we intend to adopt a non-flat kernel function, which gives more weight in the inner region rather than using equal weight everywhere. Such a non-flat kernel function assists in locating halo centres more robustly, and this is particularly important in identifying subhalos. Comparing to field halos in cosmological simulations, subhalos usually reside in more complicated density environments. It is more robust to locate subhalo centre when giving more weight to the central region. On the contrary, using a flat kernel fails to identify some subhalos; see Appendix \ref{ap:kernel} for a quantitative comparison between using flat and non-flat kernel functions to identify subhalos.

To construct a non-flat kernel function, we start from the spherically symmetric Plummer density profile, which was firstly introduced by \citet{Plummer1911} to describe the stellar distribution of globular clusters,
\begin{equation}
 \rho(r)=\frac{3M}{4\pi r_s^3} \frac{1}{(1+ r^2/r_s^2)^{5/2}},
\end{equation}
where $M$ is the total mass and $r_s$ is the scale radius. The corresponding Plummer potential is
\begin{equation}
 \phi(r)=-\frac{GM}{(r^2+r_s^2)^{1/2}},
\end{equation}
where $G$ is the gravitation constant, and the corresponding force is
\begin{equation}
 \bm{F}(\bm{r})=-\frac{GM\bm{r}}{(r^2+r_s^2)^{3/2}}.
\end{equation}
Our Plummer kernel function is constructed as the following form,
\begin{equation}
 K_{p}(r;r_s)=\frac{C}{(r^2+r_s^2)^{3/2}},
\end{equation}
where $C$ is a normalization constant, while $r$ and $r_s$ are respectively corresponding to $\left| \bm{x}_i-\bm{x} \right|$ and $b$ in Equation \ref{eq:mean_shift_vector}.  In this work, we set $r_s = 3\epsilon$ with $\epsilon$ being the gravitational softening length from simulations \citep{Dyer1993}. As If a particle is infinitely close to a Plummer potential minimum, then the Plummer style mean-shift vector 
\begin{equation}
    \bm{M}_b^{K_p} \propto\sum_{i=1}^{N}\frac{\bm{r}_i}{(r_i^2+r_s^2)^{3/2}}
\end{equation}
is proportional to the sum of Plummer force. With this kernel function, shifting from an initial seed in iterations approximately follows the force exerted by the corresponding Plummer potential. In other words, the iterative procedure mimics the process of a particle falling into a Plummer potential well. Note, BDM halo finder \citep{Klypin1997, Riebe2013} adopts a scheme essentially identical to mean-shift algorithm with a flat kernel to locate density peaks, while it was called ``sphere jittering" mechanism in the paper.

\subsection{Seeds Function}\label{subsec:seeds_func}

A seeds function is necessary to generate the initial positions (i.e. seeds) to start the kernel-shift algorithm. The simplest scheme for seeds generation is to randomly choose positions of a certain fraction of simulation particles. However, some small targets in low-density regions may be missed accidentally in such a scheme, and thus additional seeds should be placed in these low-density regions \citep{Klypin1997}. Also note that the number of seeds has direct impact on the amount of computational expense, i.e. more seeds causes more subsequent calculations (or computation time), whilst they do not affect the final results on the number of peaks. For an optimal seeds function, we expect that every candidate halo/subhalo should contain at least one seed in order to avoid missing, but in the meantime, the total number of seeds should be as few as possible in order to reduce computational cost. 

In HIKER, we develop a novel approach to generate seeds as detailed below. Firstly, for each particle in the simulation, we search its 20 nearest neighbor particles using k-d trees \citep{Bentley1975}, and compute its distance to the 20th nearest particle, $d_{20}$, which is an approximate proxy of its local density. Secondly, we use a predefined distance threshold, $r_c$, which is similar to the linking length parameter in an FOF algorithm but is slightly larger to be more conservative, to exclude those particles whose local densities are too low to be associated with any halo. We set $r_c$ as $0.24$ times the mean inter-particle separation in our code. Third, we loop over the remaining particles and adopt the particles whose $d_{20}$ are the smallest among the $d_{20}$ of their $20$ nearest neighbor particles (i.e. those particles which correspond to the local density maxima) as candidate seeds. In the final step, note that the candidate peaks which are close to each other are still possible to be moved to the same halo centre during the kernel-shift iteration. Therefore, to further reduce and computational cost, we loop over the candidate seeds, and for each candidate seed, we search its neighbouring candidate seeds within a spherical region with a radius defined as 3 times of the scale radius of the Plummer kernel function (see Section \ref{subsec:mean_shift} for details). A candidate seed peak will be determined as a true seed if its local density is still the maximum among its neighbors. Note, in this procedure, the seeds we identified fairly represent the realistic density peaks. Starting from each seed, we select all particles inside a radius of $r_K = 10 \times \epsilon$ and apply for mean-shift procedure to derive all density peaks. Note, more than one seeds may converge to a singel peak with mean shift algorithm.

\subsection{halo/subhalo identification}
Based on density peaks, we follow an approach of SO halo finder to identify field halo \citep{Lacey1994}.  The density peaks are sorted in descending order according to their local densities which are estimated from the distance to the 20th nearest neighbour. We firstly consider each peak as the centre of a field halo, while many of them are indeed centres of subhalos.   Starting from the density peak with the highest local density, then for each peak, a sphere is grown around centre until the enclosed mean density is equal to 200 times the critical density. Then we further distinguish halo and subhalo using simple geometrical relation, namely if a halo is contained in the other larger halo, the smaller halo is labeled as a subhalo. Once halos/subhalos are distinguished, we need to define subhalo boundary because only subhalo halo centres are known at the stage. There are various definitions of subhalo boundary in the literature. For examples, SUBFIND uses saddle points, HBT \citep{Han2012,Han2018} groups the bound particles of its progenitor as a subhalo. In HIKER we use a physically motivated value--tidal radius to define the boundary a subhalo. In practise, starting from each subhalo centre, a sphere is grown until its tidal radius reaches.

A halo/subhalo is usually regarded as a gravitationally bound structure, and thus it is often for some halo finders to remove unbound particles from the candidate halos/subhalos (i.e. unbinding procedure). In HIKER, we carry out the unbinding procedure following that of the AHF halo finder except of determining the bulk velocity for a potential halo/subhalo. For that we use our Plummer kernel to weight particle velocities according to their distances to the centre, instead of using a small fraction of particles within the central region to get the mean velocity as the bulk velocity. For other details of the unbinding procedure, we refer the reader to \cite{Knollmann2009}.

\section{Test results}\label{sec:results} 

In this section, we test our halo finder in three aspects using three different types of simulation data, i.e. we use (i) a set of mock halos to test the accuracy of halo properties given by HIKER  (Section \ref{subsec:mock_halo_tests}), (ii) a suite of large-scale full-box cosmological simulations to test the ability in identification of field halos (Section \ref{subsec:halo_tests}), (iii) the data from the Aquarius project \citep{Springel2008} to test the ability to identify  subhalos (Section \ref{subsec:subhalo_tests}). Note that these data have been used in the halo/subhalo finder comparison \citet{Knebe2011} and \citet{Onions2012}, and thus almost all our results can be directly compared to those of two studies. For the tests on mock halos, we will compare the HIKER results with those from all halo finders discussed in \citet{Knebe2011}, while for the other two tests, to be concise, we mainly compare the HIKER results with two widely used halo finders, SUBFIND and AHF.

\subsection{Mock halos}\label{subsec:mock_halo_tests}
Mock halos, whose properties are known by construction, are introduced in \citet{Knebe2011} to examine the accuracy of different halo finders in recovering halo/subhalo properties. Two sets of mock halo data have been prepared with a given NFW density profile and a given Plummer density profile, respectively. For each mock set, three setups have been generated: (i) an isolated host halo only; (ii) an isolated host halo + a subhalo at $0.5 R_{100}^{\rm host}$; (iii) an isolated host halo + a subhalo at $0.5 R_{100}^{\rm host}$ + a subsubhalo at $(0.5 R_{100}^{\rm host} + 0.5 R_{100}^{\rm subhalo})$. Here, $R_{100}^{\rm host}$ ($R_{100}^{\rm subhalo}$) is the radius of the host halo (subhalo) within which the mean density is $100$ times the critical density. The properties of mock halos are summarized in Table \ref{Tab:mockhalo}; see \citet{Knebe2011} for further details of these mock data.

\begin{table}
\begin{center}
\caption{The (sub)halo properties of the NFW and Plummer mock data. $N_{\rm 200}$ and $M_{\rm 200}$ are the particle number and virial mass inside the virial radius, $R_{200}$. $R_{\rm s}$ is the scale radius of the corresponding profile, and $V_{\rm max}$ is the maximum of the rotation curve.}
\label{Tab:mockhalo}

\begin{tabular}{llrcccc}
\hline
Profile    & Type &  $N_{\rm 200}$ & $M_{\rm 200}[h^{-1}\mathrm{M}_{\sun}]$ & $R_{\rm 200}[h^{-1}\mathrm{kpc}]$ & $R_{\rm s}[h^{-1}\mathrm{kpc}]$ & $V_{\rm max}[\mathrm{km} \: \mathrm{s}^{-1}]$ \\
\hline
NFW            & host      & 760,892 & 7.61 $\times 10^{13}$ & 689.1 & 189.5 & 715\\
               & sub       &  8,066  & 8.07 $\times 10^{11}$ & 151.4 & 17.0  & 182\\
               & subsub    &  84    & 8.42 $\times 10^{9}$  &  33.1 & 2.6   & 43\\
\hline
Plummer        & host      & 966,326 & 9.66 $\times 10^{13}$ & 760.5 & 190.0 & 961\\
               & sub       &  9,937  & 9.94 $\times 10^{11}$ & 161.7 & 17.0  & 314\\
               & subsub    &  100   & 10.00$\times 10^{9}$  & 23.9  & 2.6   & 79\\
\hline
\end{tabular}
\end{center}
\end{table}

In the following, we will use these two sets of mock halos to test the accuracy of HIKER in recovering halo/subhalo properties. Especially, as the HIKER method is closely related to the Plummer model, it is interesting to see how accurately it can recover the properties of halos/subhalos with Plummer and non-Plummer density profiles.

We use HIKER with the same parameter setup to identify halos/subhalos in the NFW and Plummer mocks, except of the bandwidth parameter $b$. For the Plummer mocks, $b$ is set to be the scale radius fitted directly from the data with a Plummer profile, while it is set to be a fixed value  $3\epsilon$ for the NFW mocks.

\begin{figure}[htbp]
\centering
\includegraphics[width=0.8\textwidth]{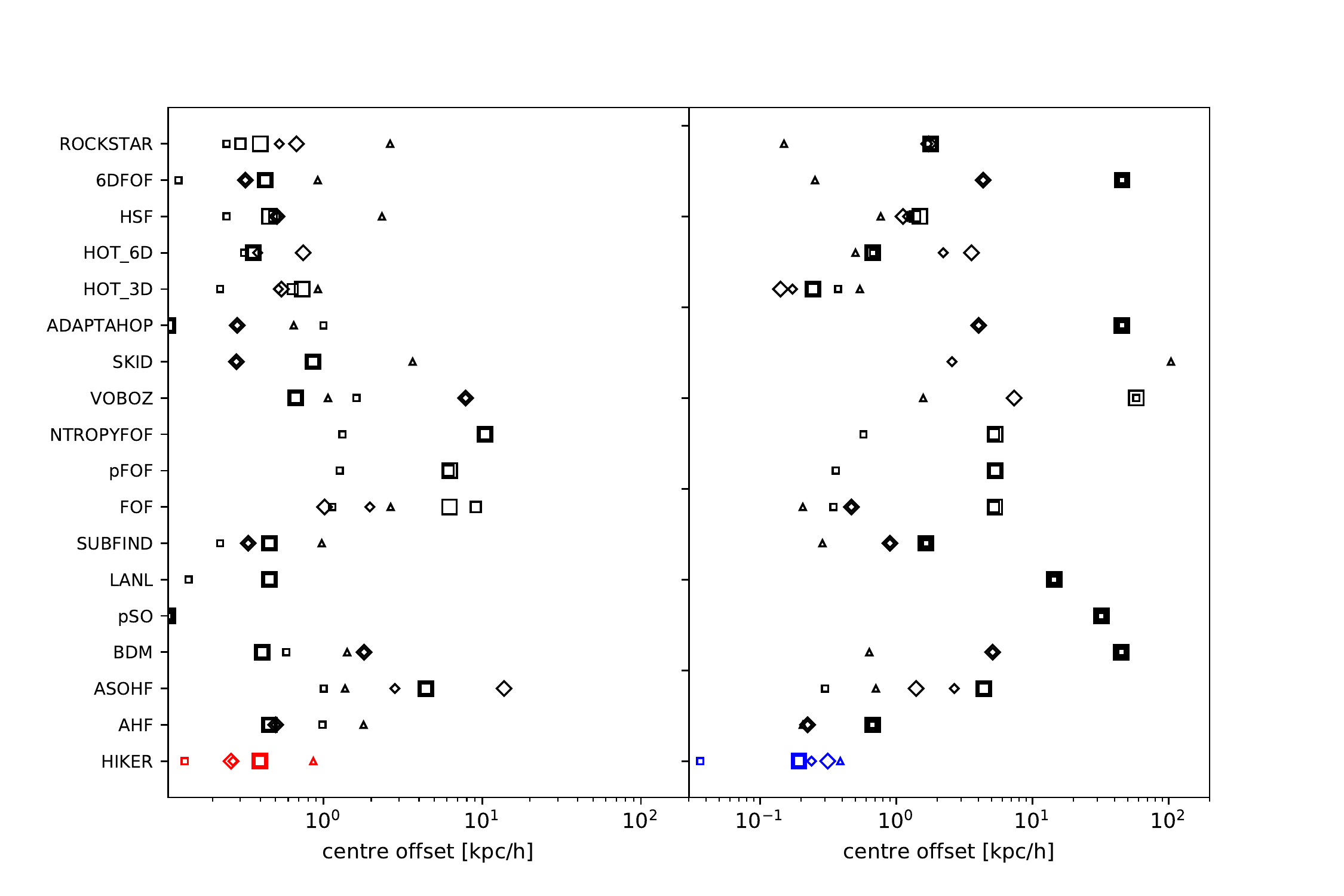}
\caption{Center offset between the actual center and the value recovered by HIKER. The left panel plots the results for NFW mock halos, while the right panel plots the results of Plummer mock halos. The results of different halo finders from \citet{Knebe2011} are plotted with black color at different $y$-coordinates, and the corresponding names are labeled on the $y$-axis. The HIKER results are highlighted with red and blue colors. In both panels, host halos, subhalos, and subsubhalos are marked with squares, diamonds, and triangles, respectively. The symbol size distinguishes the results of different setups, with larger symbols representing setups containing more substructures. Specifically, taking the ROCKSTAR results on the left panel as an example, the three squares from left to right show the results of host halos from setup (i), (ii), and (iii) respectively, the two diamonds from left to right show the subhalos from setup (ii) and (iii) respectively, and the triangle shows the subsubhalo from setup (iii).}
\label{fig:centre_offset}
\end{figure}

Following \citet{Knebe2011}, we first quantify the accuracy of HIKER in recovering halo/subhalo centres by computing the distance offset between the centre returned by HIKER and the actual one. Results are shown with red and blue symbols in Fig. \ref{fig:centre_offset}, in which the results of other 17 halo finders\footnote{These halo finders are AHF \citep{Knollmann2009}, ASOHF \citep{Planelles2010},  BDM \citep{Klypin1997}, pSO \citep{Sutter2010}, LANL \citep{Habib2009}, SUBFIND \citep{Springel2001}, FOF, pFOF \citep{Courtin2011, Rasera2010}, NTROPYFOF \citep{Gardner2007, Gardner2007a}, VOBOZ \citep{Neyrinck2005}, SKID \citep{Stadel2001}, ADAPTAHOP \citep{Aubert2004, Tweed2009}, HOT\_3D, HOT\_6D \citep{Ascasibar2005, Ascasibar2010}, HSF \citep{Maciejewski2009}, 6DFOF \citep{Diemand2006} and ROCKSTAR \citep{Behroozi2013}.} from \citet{Knebe2011} have also been shown with black symbols for easy comparison. We use squares, diamonds, and triangles to plot the centre offsets of host halos, subhalos, and subsubhalos respectively. The symbol size distinguishes different mocks, with larger symbol corresponding to the mocks with more subhalos.

For NFW mocks (left panel of Fig. \ref{fig:centre_offset}), HIKER recovers the input halo properties fairly well. Especially for the isolated host halo of setup (i), the halo centre recovered by HIKER only deviates $0.13$ $h^{-1}\mathrm{kpc}$ (i.e. $\sim 2 \times 10^{-4} R_{200}$) from the actual centre. For host halos containing substructures, due to the asymmetry in the density field caused by nesting substructure, the deviations become slightly larger ($\sim 0.4$ $h^{-1}\mathrm{kpc}$, or $\sim 6 \times 10^{-4} R_{200}$), but they are still smaller than the centre offsets of most halo finders. For subhalos and subsubhalos, HIKER also recovers their centres more accurately than many other halo finders.

As we can see from the right panel of Fig. \ref{fig:centre_offset}, HIKER performs even better for Plummer mocks. For the isolated host halo, the deviation from the input halo centre is less than $0.04$ $h^{-1}\mathrm{kpc}$ ($\sim 5 \times 10^{-5} R_{200}$), much smaller than those from other halo finders. For the halos, subhalos, and subsubhalos in setups (ii) and (iii), HIKER also performs better than other halo finders. This may be a natural outcome of HIKER which adopts a Plummer kernel function to locate density peaks.  

\begin{figure}[htbp]
\centering
\includegraphics[width=0.8\textwidth]{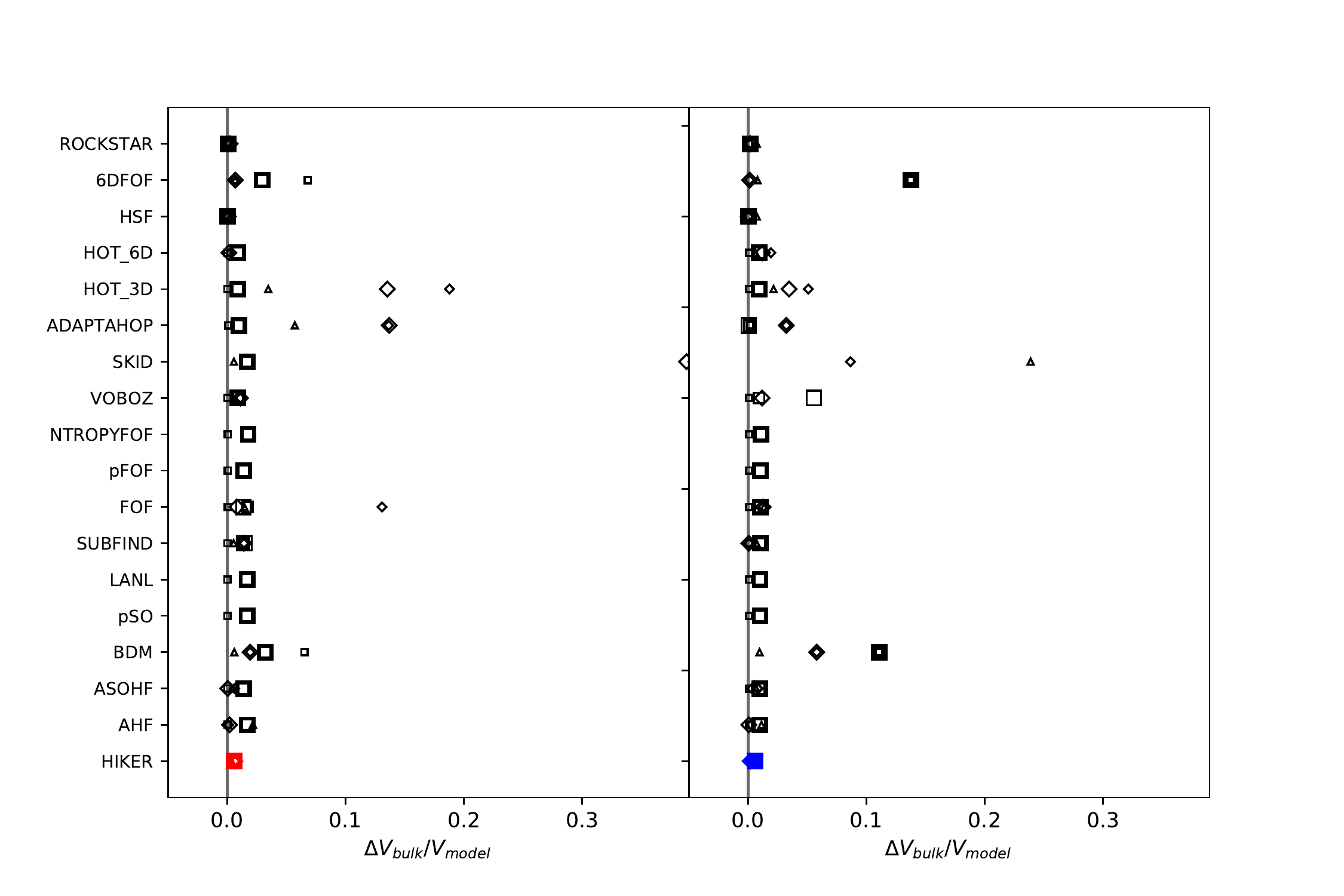}
\caption{Fractional differences between the input halo bulk velocity in mock data and the recovered bulk velocities from different halo finders. The grey vertical lines in both panels indicate no difference from the model analytical value. The layout and the symbols are in accordance with Fig. \ref{fig:centre_offset}.}
\label{fig:vbulk}
\end{figure}

\begin{figure}[htbp]
\centering
\includegraphics[width=0.8\textwidth]{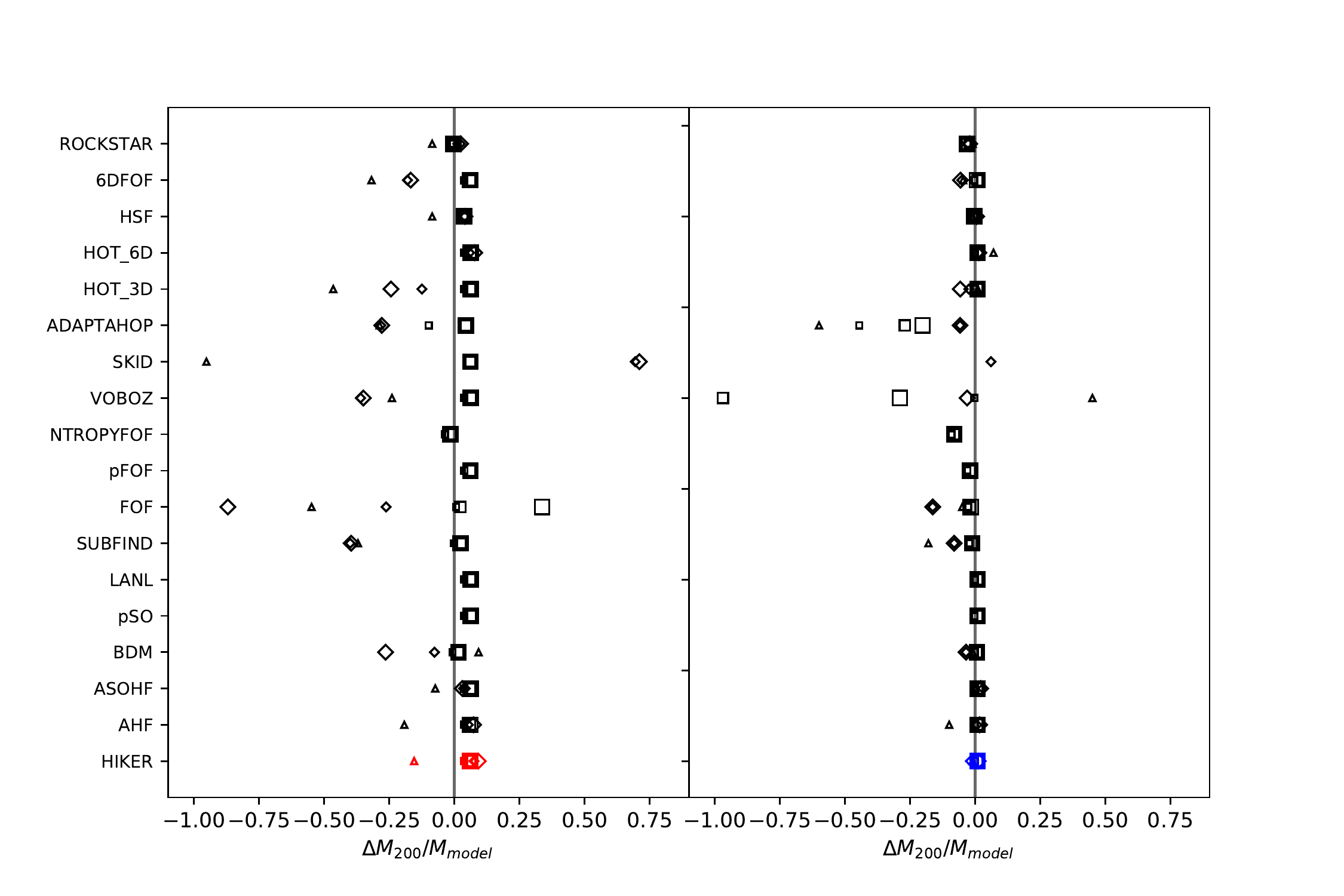}
\caption{Similar to Fig. \ref{fig:vbulk}, but for the virial mass.}
\label{fig:m200}
\end{figure}

\begin{figure}[htbp]
\centering
\includegraphics[width=0.8\textwidth]{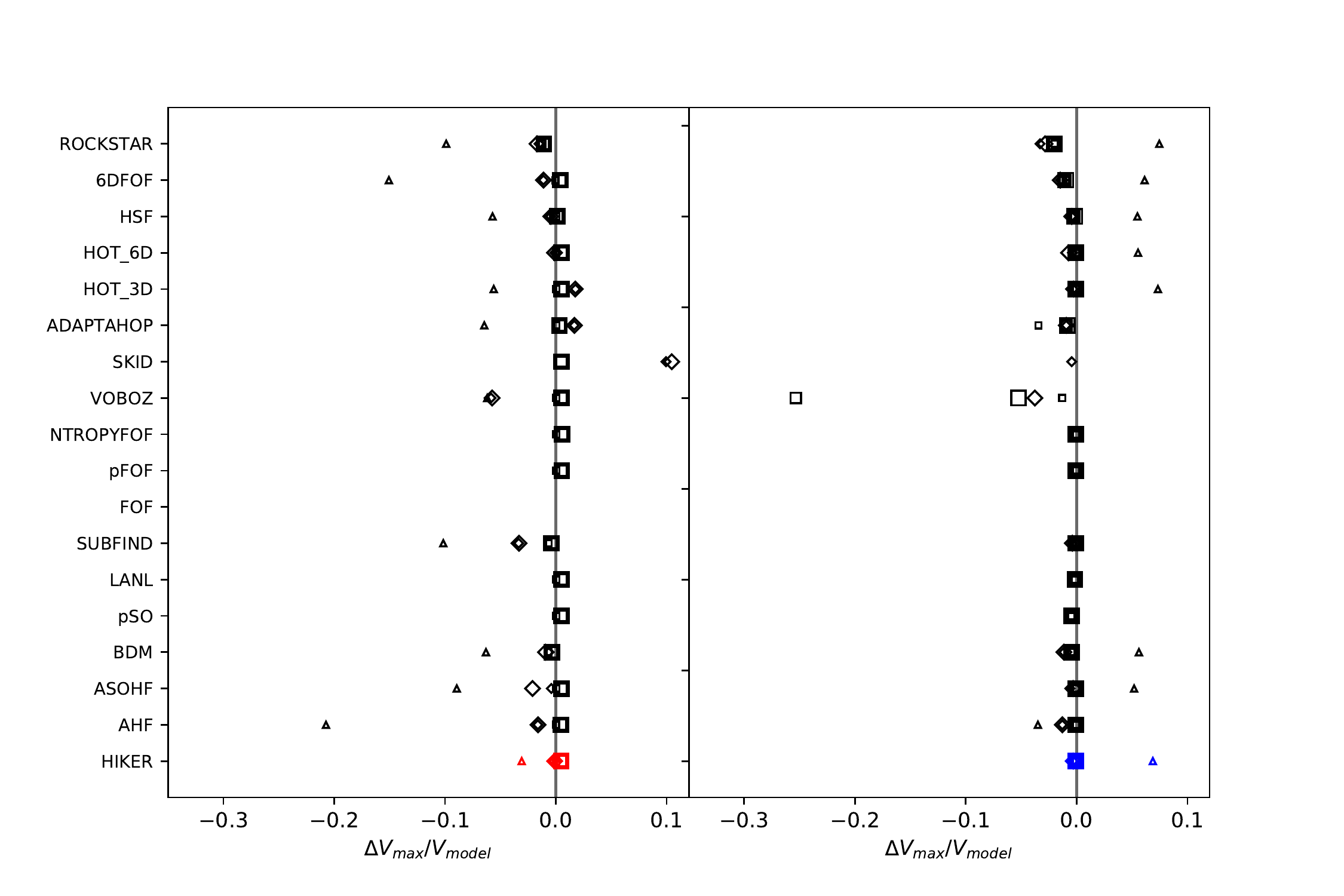}
\caption{Similar to Fig. \ref{fig:vbulk}, but for the maximum circular velocity.}
\label{fig:vmax}
\end{figure}

We also investigate how accurate HIKER can recover some other halo properties, including bulk velocities, virial masses, and the maximum circular velocities for both halos and subhalos (i.e. $x = V_\mathrm{bulk}, M_{200}$, and $V_\mathrm{max}$), and the results are presented in Figs \ref{fig:vbulk}, \ref{fig:m200} and \ref{fig:vmax} respectively. Note, the accuracy is defined as the fractional difference,
\begin{equation}
    \frac{\Delta x}{x_\mathrm{model}} \equiv \frac{x_\mathrm{code} - x_\mathrm{model}}{x_\mathrm{model}},
\end{equation}
where $x_\mathrm{code}$ and $x_\mathrm{model}$ stand for the halo properties computed by a halo finder and the input mock properties respectively. The layout and symbols in these figures are similar as those in Fig. \ref{fig:centre_offset}.

From Figs \ref{fig:vbulk}, \ref{fig:m200} and \ref{fig:vmax}, it is easy to find that HIKER recovers the aforementioned properties quite successfully both for field halos and subhalos (i.e. the fractional differences are less than $1\%$ for $V_\mathrm{bulk}$ and $V_\mathrm{max}$, and less than $6\%$ for $M_{200}$), and the HIKER results are usually better than (or comparable to) those of the other halo finders. For subsubhalos, the HIKER recoveries of $M_{200}$ (i.e. $\sim 15\%$ for NFW mocks) and $V_\mathrm{max}$ (i.e. $\sim 3\%$ for NFW mocks and $\sim 7\%$ for Plummer mocks) are not so good as the cases for field halos and subhalos, but it is still better than many other halo finders. We also find that, similar to the result for centre offsets, HIKER generally recovers the halo properties better for Plummer mocks than for NFW mocks.

As mentioned before, both BDM and HIKER are based on the mean-shift algorithm but with different kernel functions. Comparing the results between BDM and HIKER in Figs \ref{fig:centre_offset}-\ref{fig:vmax}, HIKER recovers the halo properties better than BDM which uses a flat kernel function. This is a consequence a non-flat kernel adopted in HIKER.

From the discussions in this subsection, we conclude that HIKER is quite successful in recovering halo properties. 

\subsection{Field halos}\label{subsec:halo_tests}
In this subsection, we use a suite of large-volume cosmological simulations to test the accuracy of HIKER in identifying field halos. The simulation data comes from the `Haloes gone MAD' halo-finder comparison project \citep{Knebe2011}, and it consists of three simulations with different mass resolutions, i.e. containing $256^3, 512^3$, and $1024^3$ dark matter particles respectively. These simulations have simulated the formation and evolution of the large-scale structures with the GADGET2 code \citep{Springel2005} in a comoving periodic box with a size $500$ $h^{-1}\mathrm{Mpc}$ on a side. The adopted cosmological parameters are $\Omega_m =0.3, \Omega_b=0.045, \Omega_\Lambda=0.7$, and $h=0.7$, respectively. The simulation names, dark matter particle masses, and softening lengths are summarized in Table \ref{Tab:MAD-Halo}. For each simulation, we only use the snapshot at $z=0$ to perform our tests.

\begin{table}
\begin{center}
\caption{Details of the large-volume cosmological simulations. $N_{p, \mathrm{dm}}$ gives the total number of dark matter particles in each simulation, $m_\mathrm{dm}$ is the original uncorrected mass for dark matter particles, and $\epsilon$ is the comoving softening length.}
\label{Tab:MAD-Halo}

\begin{tabular}{ccccc}
\hline
Names    & $N_{p, \mathrm{dm}}$ &  $m_{\rm dm}[h^{-1}\mathrm{M}_{\sun} ]$ & $\epsilon[h^{-1}\mathrm{kpc}]$ \\
\hline
MAD-Halo-256            & $256^3$     & 5.40 $\times 10^{11}$     & 50    \\
MAD-Halo-512            & $512^3$     & 6.59 $\times 10^{10}$     & 25    \\
MAD-Halo-1024           & $1024^3$    & 8.24 $\times 10^{9}$      & 15    \\

\hline
\end{tabular}
\end{center}
\end{table}

Note that the original simulations contain both dark matter and gas particles. However, following \citet{Knebe2011}, we only use dark matter particles to perform identify halos, and the mass of dark matter particles in each simulation (i.e. $m_\mathrm{dm}$ shown in Table \ref{Tab:MAD-Halo}) has been scaled by multiplying a factor of $\Omega_m / (\Omega_m - \Omega_b)$ accordingly.

\begin{figure}[htbp]
\centering
\includegraphics[width=1\textwidth]{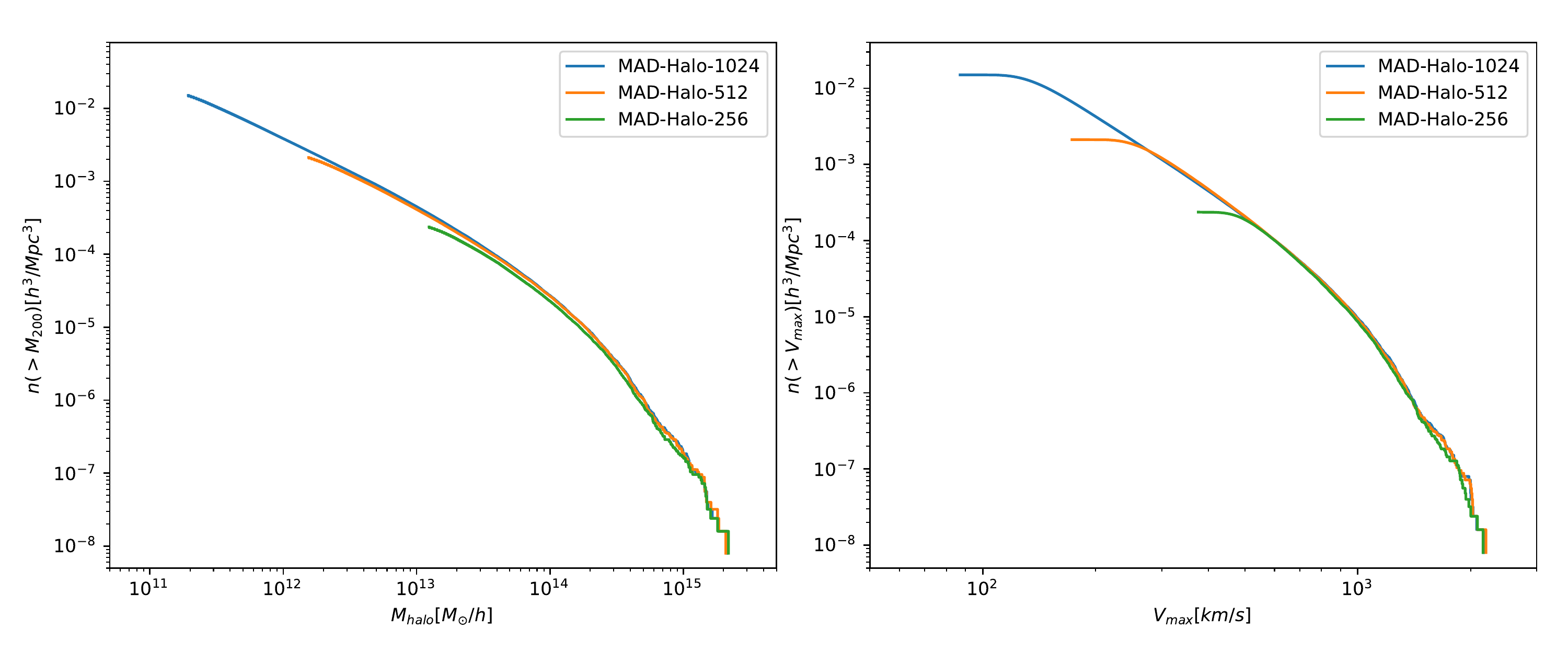}
\caption{Cumulative mass functions (left) and $V_\mathrm{max}$ functions (right) computed from HIKER field halo catalogues for three simulations with different mass resolutions.}
\label{fig:convergence_field}
\end{figure}

We identify field halos containing at least $20$ particles from all three simulations. Their cumulative mass functions and $V_\mathrm{max}$ functions are present in Fig. \ref{fig:convergence_field}. As these three simulations have the same phases in the initial conditions and they only differ in resolutions, we expect that the halo mass functions as well as $V_\mathrm{max}$ functions should converge in the reliable mass range among different resolution simulations. This is indeed true in Fig. \ref{fig:convergence_field}, indicating that HIKER works successfully in simulations with different mass resolutions.

\begin{figure}[htbp]
\centering
\includegraphics[width=1\textwidth]{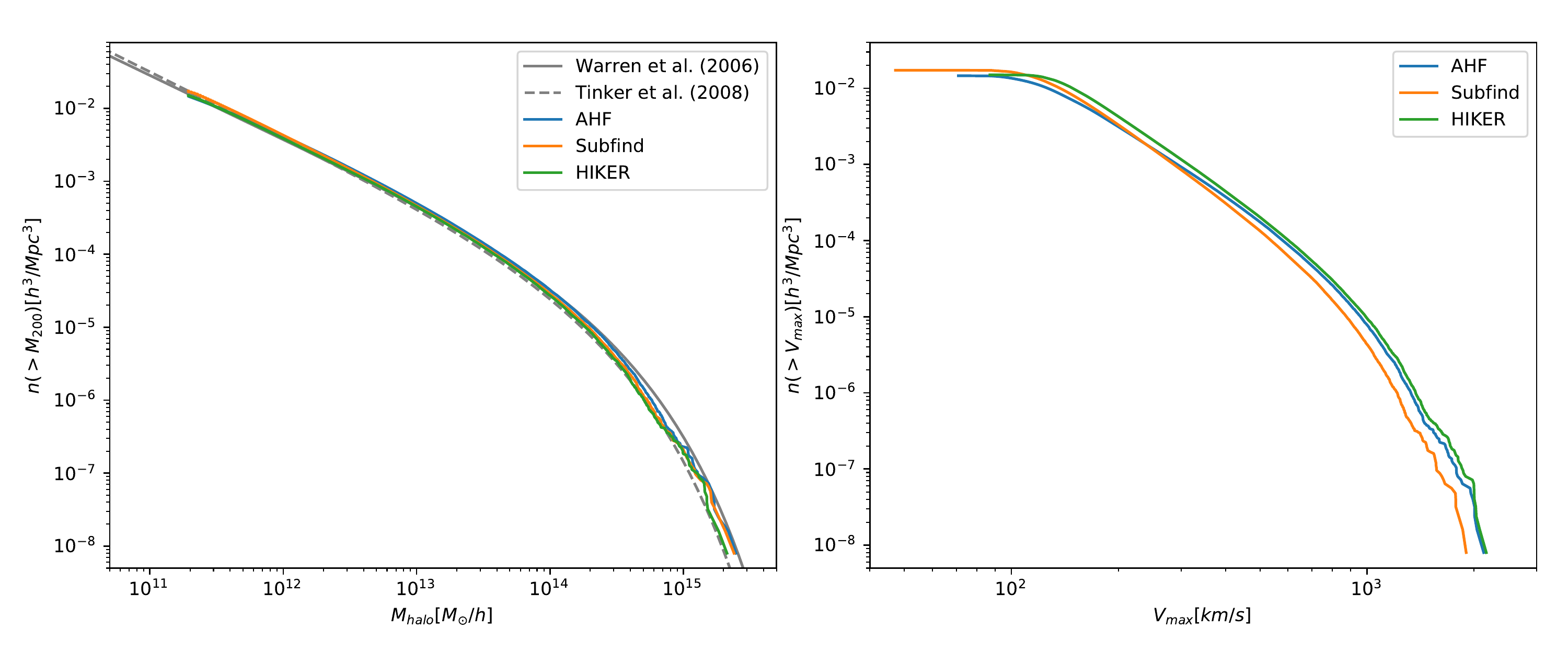}
\caption{Cumulative mass functions (left) and $V_\mathrm{max}$ functions (right) functions from the MAD-Halo-1024 simulation. The results from AHF, SUBFIND, and HIKER are plotted with blue, orange, and green lines, respectively. In the left panel, for comparisons, we over-plot the mass functions from \cite{Warren2006} and \cite{Tinker2008} with grey solid and grey dashed lines respectively.}
\label{fig:mass_vmax_field}
\end{figure}

We then compare the halo mass function and $V_\mathrm{max}$ function obtained from HIKER in the MAD-Halo-1024 simulation with those from SUBFIND and AHF in Fig. \ref{fig:mass_vmax_field}, here the SUBFIND and AHF results come from \citet{Knebe2011}, and we refer the reader to Figs 17 and 18 of the paper for more results of other halo finders. We also over-plot the analytical halo mass functions as given by \citet{Warren2006} and \citet{Tinker2008} in the figure for comparison.

We can see that the HIKER mass function agrees with the SUBFIND and AHF very well in all mass range. In the reliable mass range, all three mass functions from halo finders lie between the parameterized mass functions of \citet{Warren2006} and \citet{Tinker2008}. For the $V_\mathrm{max}$ function, HIKER tends to be slightly higher than the other two, and it is more evident in the lower $V_\mathrm{max}$ end. As there is no such difference in the halo mass functions among these halo finders, this implies that some HIKER halos (especially some with lower masses) tend to have higher $V_\mathrm{max}$, or equivalently deeper inner potentials, comparing to AHF or SUBFIND results. This possibly comes from the fact that HIKER locate halo centres more accurately, resulting in larger $V_\mathrm{max}$ in low mass halos.

\subsection{Subhalos}\label{subsec:subhalo_tests}
We use the Aq-A halo from the Aquarius project  \citep[see][for details]{Springel2008} to test HIKER in identifying subhalos. The Aq-A halo has been re-simulated with five different resolutions,  here we use three of them, i.e. Aq-A-4, Aq-A-3, and Aq-A-2, to perform our tests. Among these three simulations whose details are summarized in Table \ref{Tab:Aq_data}, the Aq-A-2 has the highest mass resolution while the Aq-A-4 has the lowest mass resolution. For all three simulations, we select a cubic region with edge length of $1$ $h^{-1}\mathrm{Mpc}$ centring the  the position of $\bm{r}_\mathrm{fiducial} = (57060.4, 52618.6, 48704.8) \: h^{-1}\mathrm{kpc}$ which is the fiducial centre defined in \citet{Onions2012} to run our halo finder. Note that within this selected region, the number of low-resolution particles is extremely few (i.e. less than $10$), and thus we simply leave out these low-resolution particles when running HIKER. In the following, we mainly compare the HIKER results with the AHF and SUBFIND ones, and the reader can refer to \citet{Onions2012} for results of other halo finders.

\begin{table}
\begin{center}
\caption{Some details of the Aq-A halos used in this study. $N_\mathrm{hres}$ ($N_\mathrm{lres}$) is the number of high-resolution (low-resolution) particles in the simulation, $N_\mathrm{select}$ is the number of high-resolution particles within our selected region (i.e. a cubic region with edge length of $1$ $h^{-1}\mathrm{Mpc}$ centring $\bm{r}_\mathrm{fiducial} = (57060.4, 52618.6, 48704.8) \: h^{-1}\mathrm{kpc}$), $m_{p, \mathrm{hres}}$ is the mass of high-resolution particles, and $\epsilon$ is the comoving softening length.}
\label{Tab:Aq_data}
\begin{tabular}{lrrrrr}
\hline
Name & $N_\mathrm{hres}$ & $N_\mathrm{lres}$ & $N_\mathrm{select}$ & $m_{p,\mathrm{hres}}[h^{-1}\mathrm{M}_{\sun}]$ & $\epsilon[\mathrm{pc}]$ \\
\hline
Aq-A-4 & 18,535,972    & 634,793     &7,434,975    & $2.868\times10^{5}$ & 342.5 \\
Aq-A-3 & 148,285,000   & 20,035,279  &59,347,132   & $3.585\times10^{4}$ & 120.5 \\
Aq-A-2 & 531,570,000   & 75,296,170  &212,792,272  & $1.000\times10^{4}$ & 65.8 \\
\hline
\end{tabular}
\end{center}
\end{table}

\begin{figure}[htbp]
\centering
\includegraphics[width=1\textwidth]{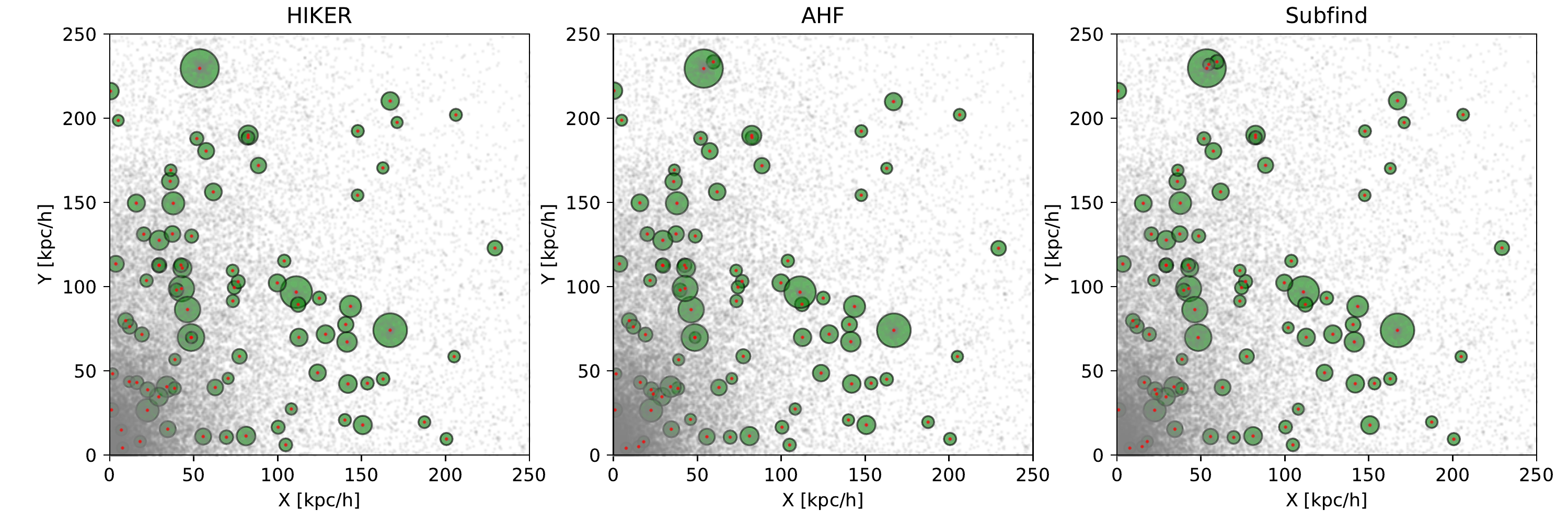}
\caption{Visualization of subhalo-finding results from HIKER (left), AHF (middle) and SUBFIND (right) on the Aq-A-4 data. The region shown in each panel is the same quadrant as presented in \citet{Onions2012}. The identified subhalos are indicated by red dots and green circles whose radii are scale with $V_\mathrm{max}/3$. Only subhalos with $V_\mathrm{max} > 10 \: \mathrm{km} \: \mathrm{s}^{-1}$ are shown here. The grey background shows the dark matter density computed from simulation particles.}
\label{fig:visual_aqa4}
\end{figure}

\begin{figure}[htbp]
\centering
\includegraphics[width=1\textwidth]{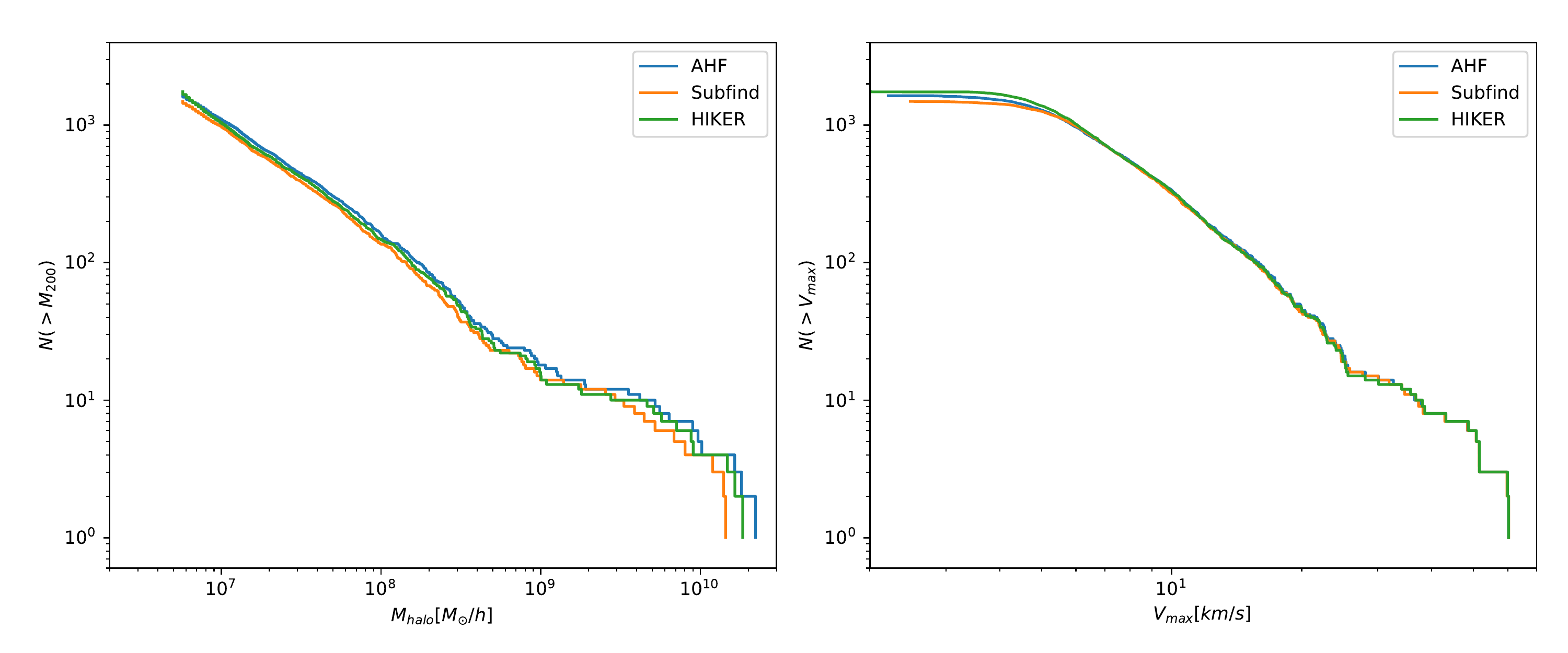}
\caption{Cumulative mass functions (left) and $V_\mathrm{max}$ functions (right) for subhalos identified from the spherical region with a radius of $250$ $h^{-1}\mathrm{kpc}$ around the fiducial position in the Aq-A-4 data. The results from AHF, SUBFIND, and HIKER are plotted with blue, orange, and green lines, respectively.}
\label{fig:mass_vmax_sub}
\end{figure}

\begin{figure}[htbp]
\centering
\includegraphics[width=1\textwidth]{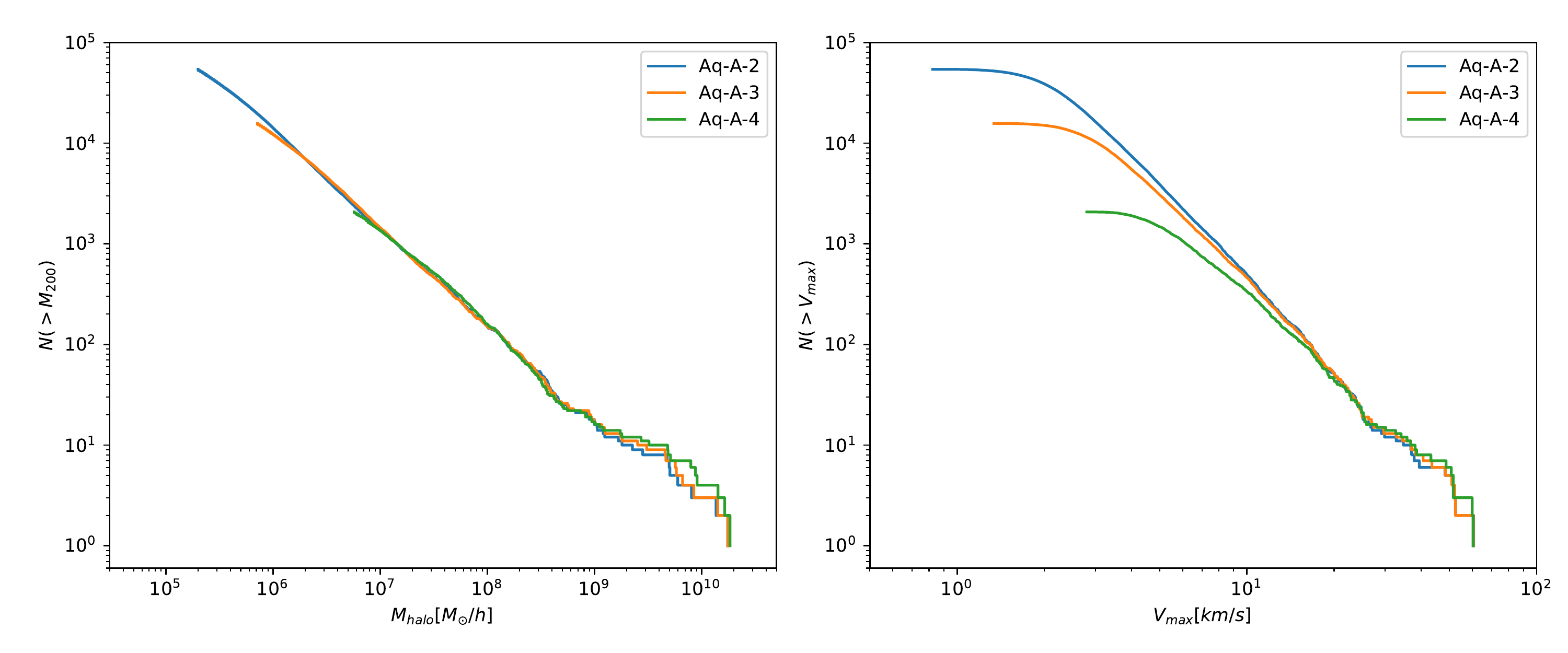}
\caption{Cumulative mass functions (left) and $V_\mathrm{max}$ functions (right) for subhalos identified from different Aq-A simulations by HIKER. Similar to Fig. \ref{fig:mass_vmax_sub}, these subhalos are from the spherical region with a radius of $250$ $h^{-1}\mathrm{kpc}$ centring the fiducial position. We use the blue, orange, and green lines to plot the results from Aq-A-2, Aq-A-3, and Aq-A-4 simulations, respectively.}
\label{fig:convergence_sub}
\end{figure}

In Fig. \ref{fig:visual_aqa4}, We first compare the subhalos identified with HIKER from the Aq-A-4 data to those identified with AHF and SUBFIND by visualisation. To be in accordance with \citet{Onions2012}, we have show the same quadrant region around the fiducial position here. Each identified subhalo is represented with a green circle whose radius scales with its $V_\mathrm{max}/3$, and the subhalo centre is marked with a red dot. Note that only subhalos with $V_\mathrm{max} > 10 \: \mathrm{km} \: \mathrm{s}^{-1}$ are plotted in this figure. Comparing to AHF and SUBFIND, HIKER misses one subhalo in the upper left corner, and it identifies a few more low-mass subhalos at the lower left corner (i.e. the region near the Aq-A halo centre). But in general, the HIKER subhalos agree very well with the AHF and SUBFIND ones in positions and $V_\mathrm{max}$.

As a quantitative comparison, in Fig. \ref{fig:mass_vmax_sub} we plot the cumulative mass functions and $V_\mathrm{max}$ functions for subhalos in the Aq-A-4 simulation identified by HIKER, AHF, and SUBFIND. The subhalos used to plot this figure are within a sphere of $250$ $h^{-1}\mathrm{kpc}$ from the fiducial position and contain at least $20$ particles. Overall the HIKER results are in line with those of AHF and SUBFIND.

We have also used HIKER to identify subhalo on the level 2 and level 3 Aq-A simulations, and the results are presented in Fig. \ref{fig:convergence_sub}. As expected, the HIKER subhalo mass functions and $V_\mathrm{max}$ functions converge very well in different resolution simulations. These results are consistent with the resolution convergence tests shown in \citet{Springel2008} with SUBFIND.

From the discussions above, we conclude that HIKER identifies subhalos with an accuracy comparable to that of the widely used AHF and SUBFIND.

\section{Conclusions and discussions}\label{sec:conclusions} 
In this work, we develop a new spherical overdensity halo/subhalo code--HIKER for cosmological simulations. HIKER employs the mean-shift algorithm combining with a Plummer kernel to efficiently and robustly locate density peaks. Based on density peaks, dark matter halos are further identified as spherical overdensity structures, and subhalos are substructures with boundaries equal to their tidal radius. We use mock halos to test our halo-finding code, and show that HIKER performs excellently in locating halo/subhalo centres and recovering halo properties. Especially, the accuracy of HIKER in recovering halo/subhalo centres is higher than most halo finders. With large-volume and zoom-in cosmological simulations, we further showed that HIKER reproduces the abundance of field halos and subhalos quite accurately, and the HIKER results are in agreement with those of two widely used halo finders, SUBFIND and AHF.

Although we only use HIKER to identify halos/subhalos from dark matter-only simulations in this study, it can be quite straightforward to extend the HIKER algorithm to include particles with different masses (e.g. gas, stars, etc.) by further multiplying the kernel function with different weights for different particle types in Equation \ref{eq:mean_shift_vector}. 

\normalem
\begin{acknowledgements}
We acknowledge support from the National Key Program for Science and Technology Research and Development (2017YFB0203300). 
SS is particularly grateful to Prof. Alexander Knebe for providing us data to do our tests as well as a lot of helpful discussions. QG and LG acknowledge support from NSFC grant (No. 11425312), and two Royal Society Newton Advanced Fellowships, as well as the hospitality of the Institute for Computational Cosmology at Durham University. QG is also supported by two NSFC grants (Nos. 11573033, 11622325), and the ``Recruitment Program of Global Youth Experts" of China, the NAOC grant (Y434011V01). 
\end{acknowledgements}

\bibliographystyle{raa}
\bibliography{ref}

\appendix
\section{Kernel effects}\label{ap:kernel}
Kernel functions are a key concept in the HIKER algorithm. To study quantitatively the effects of kernel functions on the identification of field halos, we run HIKER twice, one equipped with a flat kernel and the other with a Plummer kernel (with $b = 3\epsilon$), on the MAD-Halo-512 simulation. Similar runs are performed on the Aq-A-4 simulation data to study the effects on subhalo finding. Note that in the unbinding procedures, to estimate halo bulk velocities more reliably, in the runs with flat kernels we only use a certain fraction of particles in the central region as most halo finders do, while in the runs with Plummer kernels we utilize the Plummer kernel to give more weight on the central velocity. 

\begin{figure}[htbp]
\centering
\includegraphics[width=1\textwidth]{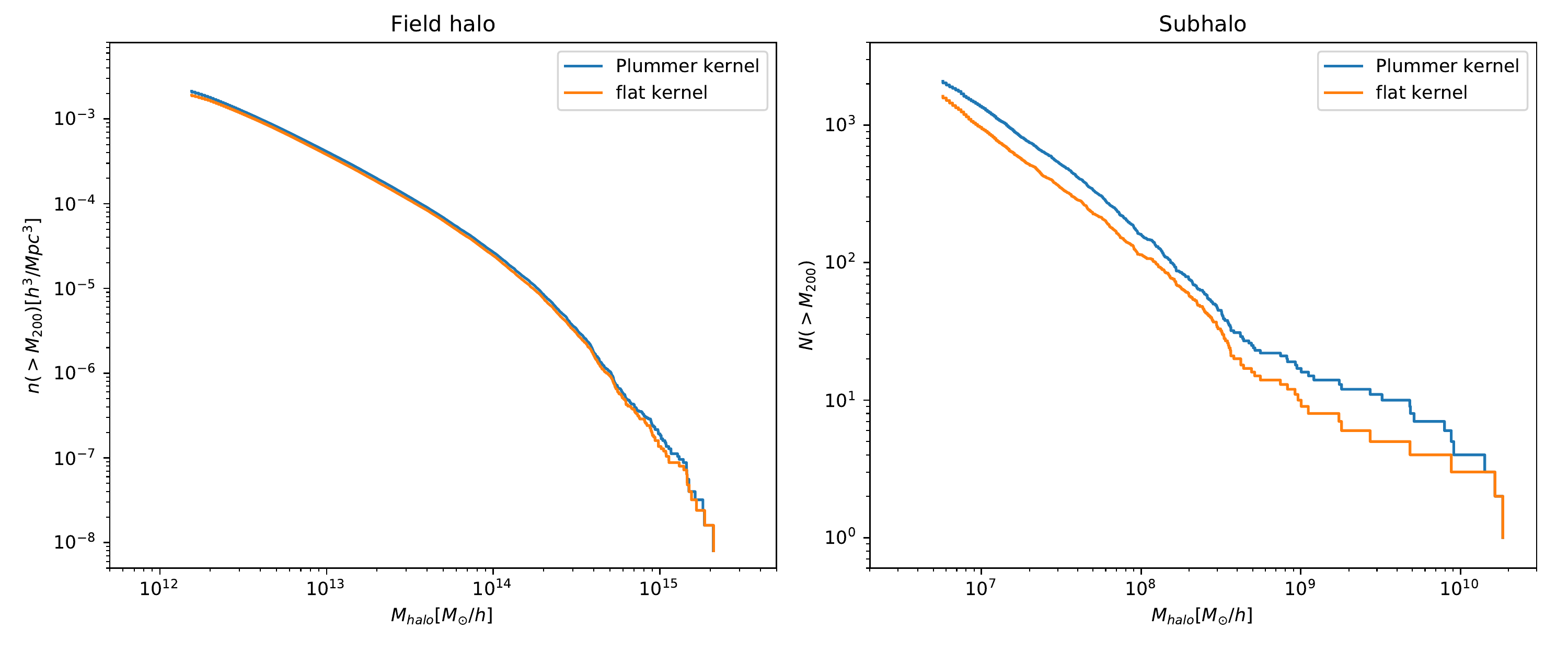}
\caption{Cumulative mass functions for field halos from MAD-Halo-512 simulation (left) and for subhalos from Aq-A-4 simulation (right). In both panels, the blue and orange lines plot the results from HIKER equipped with a Plummer kernel and a flat kernel, respectively.}
\label{fig:kernel_effect}
\end{figure}

The cumulative mass functions from these runs are summarized in Fig. \ref{fig:kernel_effect}. The field halo mass functions are barely affected by kernel functions. However, the subhalo mass functions are very sensitive to kernel functions, i.e. the number of subhalos recovered in the run with a flat kernel is much lower than that in the run with a Plummer kernel. As we have shown in Section \ref{subsec:subhalo_tests}, the HIKER subhalo results agree fairly well with those of AHF and SUBFIND. The results here point out that introducing a non-flat kernel function can help locate the halo centres in a more robust way, especially in finding subhalos. Usually, a potential subhalo is surrounded by more complex density fields, and this makes it easy for the candidate centre to shift away if the central core is not emphasized. In contrast, field halos are usually isolated, and the density environment around them is much simpler, and a flat kernel will be good enough to capture that.

As the centre locating method in BDM is equivalent to the mean-shift algorithm with a flat kernel, the results in this subsection also suggest that with a Plummer kernel function, HIKER can significantly improve BDM in identifying subhalos.

\end{document}